\documentclass[journal,transmag]{IEEEtran}

\usepackage{amsmath,amssymb,amscd,latexsym,dsfont}
\usepackage{multicol}
\usepackage{multirow}
\usepackage{float}
\usepackage{comment}
\usepackage{color}
\usepackage[caption=false, font=footnotesize]{subfig}
\usepackage{enumerate}
\usepackage{stfloats}
\usepackage{psfrag}
\usepackage{cite}
\usepackage{pgfplots}
\pgfplotsset{compat=1.5}
\usepackage[font=footnotesize]{caption}
\usepackage{filecontents}
\usetikzlibrary{arrows,automata}
\usepackage[latin1]{inputenc}
\usepackage{wrapfig}
\newcounter{lemma}

\newcommand{\vect}[1]{\mathbf{#1}}
\newcommand{\SNR}{\text{SNR}}
\newcommand{\GMI}{\text{GMI}}

\begin{document}

\title{End-to-End Learning of Geometrical Shaping\\ Maximizing Generalized Mutual Information}

\newcommand{\aff}[1]{\textsuperscript{(#1)}}
\author{
    \IEEEauthorblockN{Kadir G\"um\"u\c s\IEEEauthorrefmark{1}, Alex Alvarado\IEEEauthorrefmark{1}, Bin Chen\IEEEauthorrefmark{2}, Christian H\"ager\IEEEauthorrefmark{3}, Erik Agrell\IEEEauthorrefmark{3}}
    \IEEEauthorblockA{\\\IEEEauthorrefmark{1}Department of Electrical Engineering, Eindhoven University of Technology, Eindhoven, The Netherlands}
    \IEEEauthorblockA{\IEEEauthorrefmark{2}School of Computer Science and Information Engineering, Hefei University of Technology, Hefei, China}
     \IEEEauthorblockA{\IEEEauthorrefmark{3}Department of Electrical Engineering, Chalmers University of Technology, G\"oteborg, Sweden\\\
     k.gumus@student.tue.nl}
}

% The paper headers
\markboth{PREPRINT, DECEMBER 11, 2019}%
{Shell \MakeLowercase{\textit{et al.}}: Bare Demo of IEEEtran.cls for IEEE Transactions on Magnetics Journals}

\IEEEtitleabstractindextext{%
\begin{abstract}
GMI-based end-to-end learning is shown to be highly nonconvex. We apply gradient descent initialized with Gray-labeled APSK constellations directly to the constellation coordinates. State-of-the-art constellations in 2D and 4D are found providing reach increases up to 26\% w.r.t. to QAM.\\
\end{abstract}
}

\maketitle

\IEEEdisplaynontitleabstractindextext

\IEEEpeerreviewmaketitle

\section{Introduction}

\IEEEPARstart{S}{ignal} shaping has recently received considerable attention in the literature and is now regarded as a key technique to improve throughput in high-speed fiber-optic systems. Shaping methods can be broadly categorized into probabilistic shaping (PS) and geometric shaping (GS), both having distinct advantages and disadvantages \cite{Qu2017, Steiner2017, Millar2018}. This paper focuses on GS, i.e., using nonrectangular constellations, due to its relative simplicity compared to PS.

Traditional methods for GS include, e.g., genetic algorithms \cite{Steiner2017} and pair-wise optimizations \cite{Zhang2017a, Chen2018a}. A different approach is to regard the entire communication system design as an end-to-end reconstruction task, similar to autoencoders (AEs) in machine learning (ML) \cite{OShea2017}. This approach jointly optimizes transmitter and receiver neural networks (NNs), where the transmitter NN performs GS. A key advantage of this method is that it can be applied to arbitrary channels, including nonlinear optical ones. This was done for example in \cite{Shen2018ecoc, Jones2018, Karanov2018}, where the objective function was a lower bound on the mutual information (MI). In practice, binary forward-error correction (FEC) is typically employed, in which case the  generalized mutual information (GMI) is a more suitable performance metric. GMI-based end-to-end learning was studied in \cite{Jones2019}, where it is shown that the AE approach can jointly optimize the constellation and its corresponding binary labeling. While the optimization arrived at a well-performing solution, the irregular constellation shape in \cite[Fig.~2]{Jones2019} suggests that only a local optimum was found. 

In this paper we investigate GMI-based end-to-end AE learning systematically extending the results of \cite{Jones2019}. Our contributions are threefold.

    First, we demonstrate that the use of GMI as a cost function 
    results in a highly nonconvex optimization landscape. Since AE learning relies on local search methods (i.e., gradient descent), the optimization is thus prone to only find local optima when randomly initialized.
    Second, we compare a variety of methods to deal with nonconvex functions. Generic ML methods (e.g., cyclical learning rates \cite{Smith2017}) are shown to be relatively ineffective for our problem, whereas domain-specific methods (e.g., initialization with Gray-labeled constellations) are more promising. 

    Lastly, 
    we propose a simple, yet effective, approach to GMI-based GS which applies gradient descent directly to constellation coordinates. This is similar to the approach for symbol-error minimization proposed in 1973 in \cite{Foschini1974}. Our method can be seen as a special case of AE learning where the transmitter NN has no hidden layers. While this method is not guaranteed to converge to a global optimum, state-of-the-art constellations with up to $1024$ points in 2D and up to $64$ points in 4D are obtained and reported here. 

\section{End-to-End Autoencoder Learning Based on GMI}

We start by reviewing the AE approach in \cite{Jones2019}. Let $M$ be the constellation size and $m = \log_2 M$. At the transmitter, binary vectors $\vect{b} = (b_1, \ldots, b_m) \in \{0,1\}^m$ are mapped to constellation points $\vect{x} \in \mathbb{R}^N$ via an NN according to $\vect{x} = f_{\theta}(\vect{b})$, where $\theta$ are the NN parameters (i.e., weights and biases). The received channel observation $\vect{y} \in \mathbb{R}^N$ is passed through a receiver NN which tries to learn the bit-wise posterior distributions $f_{B_i|\vect{Y}}(b_i | \vect{y})$, $i = 1, \ldots, m$. The learned posteriors are denoted by $q_\phi(b_i | \vect{y})$, $i = 1, \ldots, m$, where $\phi$ are the parameters of the receiver NN. Training of $(\theta, \phi)$ is based on the per-sample loss $\ell(\vect{b}, \vect{y}) = - m - \sum_{i=1}^{m} \log_2 q_\phi(b_i | \vect{y})$, where the negative expected loss $-\mathbb{E}[\ell(\vect{B}, \vect{Y})]$ can be shown to be a lower bound on the $\GMI =  m + \mathbb{E}[\sum_{i=1}^m \log_2 f_{B_i|\vect{Y}}(B_i|\vect{Y})]$ \cite{Jones2019}. In practice, the expectation $\mathbb{E}[\ell(\vect{B}, \vect{Y})]$ is replaced by using empirical averages $\frac{1}{K} \sum_{i=1}^K \ell( \vect{b}^{(i)}, \vect{y}^{(i)} ) $, where $K$ is the batch size. 

Throughout this paper, we assume transmission over the $N$-dimensional additive white Gaussian noise (AWGN) channel. This channel models well uncompensated multi-span optical links with standard single-mode fiber (SSMF). In Sec.~3, reach results will be presented by assuming the Gaussian noise (GN) model \cite{PoggioliniJLT2014}. The AWGN channel assumption also allows us to focus exclusively on the transmitter optimization (i.e., GS) without having to consider the receiver optimization of $\phi$ and the associated hyperparameter tuning. In particular, in this paper we use the per-sample loss $\ell(\vect{b}, \vect{y}) = - m - \sum_{i=1}^{m} \log_2 f_{B_i | \vect{Y}} (b_i | \vect{y})$ based on exact posteriors, which can be evaluated in closed form without the need for a receiver NN since the channel law is known. 

\begin{figure*}[!t]
    \centering
    {\includegraphics{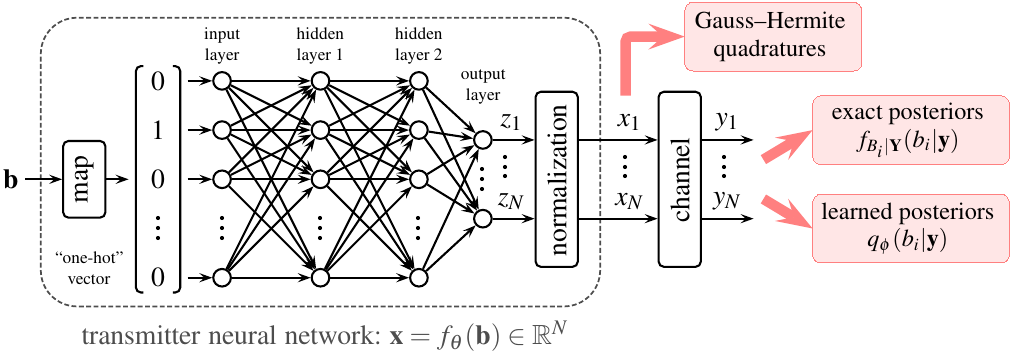}}
    $\quad$
    {\includegraphics{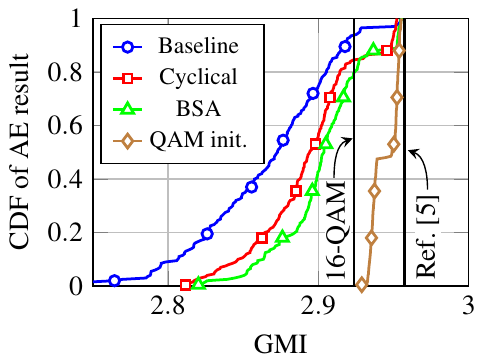}}
    %\vspace{-0.15cm}
    \caption{Left: Block diagram of the transmitter NN for GMI-based end-to-end learning; red arrows illustrate three ways to compute the loss function for optimizing the NN parameters. Right: Empirical CDF of the AE results assuming $200$ random starting points, where $M=16$, $N=2$, $\SNR = 9\,$dB (Cyclical: cyclical learning rate, BSA: binary switching algorithm, QAM init.: initialization with Gray-labeled $16$-QAM). }
    \label{fig:ae}
    % \vspace{-0.5cm}
\end{figure*} 
We implemented an AE as shown in Fig.~\ref{fig:ae} (left), where the transmitter NN has $2$ hidden layers, each with $200$ ReLU-activated neurons.\footnote{Source code and the obtained constellations can be found online at https://github.com/kadirgumus/Geometric-Constellation-Shaping} Compared to \cite{Jones2019}, we first map $\vect{b}$ to one-hot encoded vectors, rather than directly using $\vect{b}$ as the NN input (i.e., our input layer has $M$ neurons and not $m$ as in \cite{Jones2019}). The NN parameters $\theta$ are \emph{randomly initialized} using the approach in \cite{Glorot2010} and optimized using the Adam optimizer \cite{Kingma2014} with learning rate (LR) $0.001$ and batch size $K=480$. After $2000$ gradient steps, the GMI of the resulting constellation is approximated with Gauss-Hermite quadratures. This procedure is repeated $200$ times. Fig.~\ref{fig:ae} (right) shows the empirical cumulative distribution function (CDF) of the obtained GMIs (circles), where $M=16$, $N=2$, $\SNR = 9$ dB. The best constellation achieved a GMI of $2.957$ bits/2D, which is comparable to the GMI of $2.958$ bits/2D reported in \cite{Chen2018a}. However, with over $93$\% probability, the AE returned a constellation whose GMI is worse than Gray-labeled $16$-QAM. These results indicate that the optimization landscape for GMI-based learning is highly nonconvex, which makes it very challenging to find a global optimum for this problem. These results also show that the initialization is an important design parameter.

A straightforward way to deal with nonconvex functions is to repeat the optimization with more starting points. On the other hand, it would be desirable to modify the optimization in order to guarantee a better outcome. One option is to use cyclical LRs \cite{Smith2017,Loshchilov2017}, where the LR is varied between some predefined boundary values, thereby simulating multiple restarts in a single optimization run. While this indeed made the AE results more reliable (squares in Fig.~\ref{fig:ae} (right)), the chance of obtaining a constellation worse than $16$-QAM remains high. We also experimented with various NN architectures but did not notice any significant differences in terms of the optimization behavior. Besides generic ML methods, domain-specific approaches may also improve the optimization behavior. For example, the binary switching algorithm (BSA) finds the best ``swap'' of two binary labels and can overcome barriers in the optimization landscape \cite{Schreckenbach2003}. The BSA can be integrated into the AE learning by modifying the mapping function to the one-hot vectors. We repeated the simulations 
executing the BSA every $200$ gradient steps, which gave slightly more improvements compared to cyclical LRs (triangles in Fig.~\ref{fig:ae} (right)). Finally, we initialized the optimization with Gray-labeled constellations. This requires a pre-optimization step, where the NN parameters are first fitted to produce a desired constellation. Using Gray-labeled $16$-QAM as the initialization for the simulations gave the most reliable optimization outcome among all methods (diamonds in Fig.~\ref{fig:ae} (right)). Note that the optimization outcome is still random due to the nature of stochastic gradient descent and the finite batch size. 
%\vspace{-0.5em}
%\\
\section{Proposed Approach to GMI-Based Geometric Shaping}
As noted in the previous section, the particular architecture of the transmitter NN (e.g., the number of layers) appears to have little influence on the optimization behavior. On the other hand, it is always desirable to minimize the NN size and reduce the number of free parameters in order to keep the optimization time to a minimum. One extreme case is when all hidden layers are removed and the input layer is directly connected to the output layer. Assuming that all biases are $0$, the network weights then directly correspond to the coordinates of the constellation points (before the normalization). We show here that this approach is indeed sufficient to obtain state-of-the-art GMI-optimized constellations. This also connects the end-to-end AE learning approach in \cite{OShea2017} to early works on GS for symbol-error minimization, where gradient descent is directly applied to the constellation coordinates \cite{Foschini1974}.

\begin{figure*}[!t]
    \centering
    \includegraphics{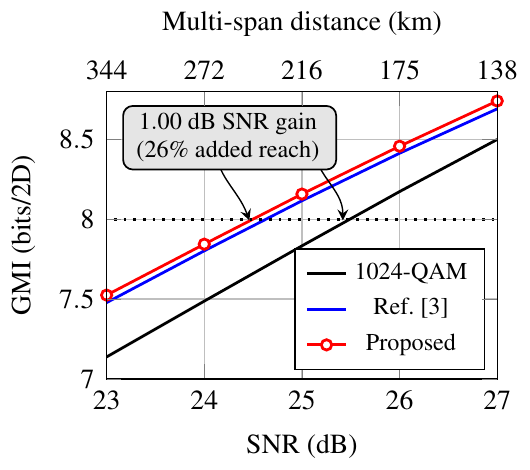}\hspace{-0.7em}\includegraphics{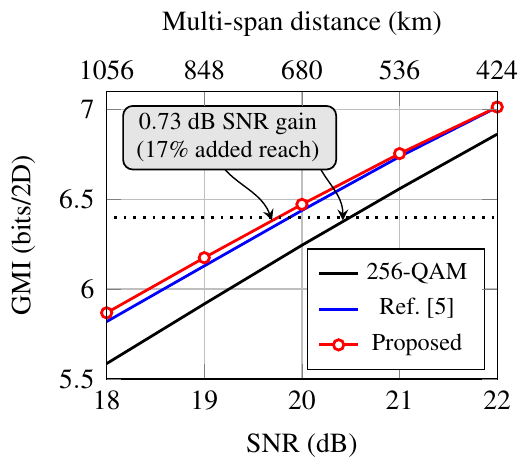}\hspace{-0.7em}\includegraphics{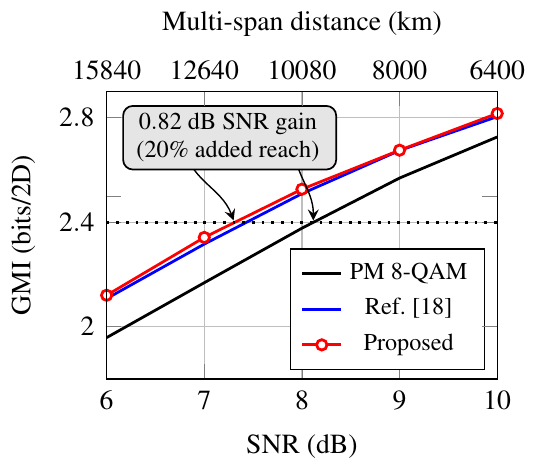}
    %\vspace{-0.4cm}
    \caption{Results for $N=2$, $M=1024$ (left), $N=2$, $M=256$  (center), and $N=4$, $M=64$ (right), where the optimization is done separately for each SNR. SNR gains are with respect to QAM and measured assuming a binary FEC with rate $0.8$ (dotted lines). The amount of rings for the initial APSK constellations are 16, 8 and 1, respectively (left to right). Gains with respect to prior works are $0.14$ dB, $0.12$ dB, and $0.13$ dB, respectively. The reach increases are calculated according to the GN model \cite{PoggioliniJLT2014} based on a 
    multi-span optical link with SSMF, 45~GBaud symbol rate,   and 11 WDM channels. The length of a span is 80 km and  EDFA noise figure is 4.5~dB.}
    \label{fig:2D}
    %\vspace{-0.5cm}
\end{figure*}

We use the simplified NN with no hidden layers, initialized with Gray-labeled APSK constellations as defined in \cite{Liu2011}. APSK initialization gave good results for medium to high SNRs. QAM initialization gave better results only at very high SNRs. In order to remove stochastic effects from the optimization, we use Gauss--Hermite quadratures to compute the GMI. Optimization results for 2D formats with $M=1024$ and  $M=256$, as well as 4D formats with $M=64$ are shown in Fig.~\ref{fig:2D}, where a separate optimization is performed for each SNR. The number of iterations was set to $1000$, $800$, and $3000$, respectively. As a comparison for $M=256$, we use the GMIs reported in \cite{Chen2018a}. We also note that the optimization procedure in \cite{Chen2018a} is much more computationally involved and requires hours to converge, whereas the proposed gradient-based approach converges within around $15$ minutes. Our proposed approach is therefore scalable also to larger constellation sizes. Indeed, for $M=1024$, no results were reported in \cite{Chen2018a}. We were unable to find GMI-optimized constellations with $M=1024$ in the literature. Instead, we implemented the approach in \cite{Millar2018} as a comparison, which also uses APSK as a starting point but only optimizes the radii distribution. For the 4D case, we compare to a recently proposed format based on polarization-ring-switching \cite{ChenJLT2019}. In all three cases, our approach gives state-of-the-art results, outperforming prior work by $0.14$ dB, $0.12$ dB, and $0.13$ dB, respectively, measured assuming a binary FEC with rate $0.8$ (dotted lines). Compared to conventional QAM formats, SNR gains of up to $1$ dB are obtained, translating into  up to $26$\% reach increases. 

\pgfplotsset{
compat=1.11,
legend image code/.code={
\draw[mark repeat=2,mark phase=2]
plot coordinates {
(0cm,0cm)
(0.2cm,0cm)       
(0.4cm,0cm)      
};
}
}

%\vspace{-0.5em}
\section{Conclusions}
%\vspace{-0.2cm}
We proposed a fast end-to-end learning algorithm to solve the problem of optimizing \emph{labeled} constellations. It was shown that the problem in question is nonconvex and that off-the-shelf algorithms are prone to converge to local optima. The crucial role of a good starting point for the optimization was also highlighted.

\vspace{0.3cm}

{%\scriptsize 
\noindent{\bf{Acknowledgements}}:
This work was funded by the European Union's Horizon 2020 research and innovation programme under the Marie Sk\l{}odowska-Curie grant no.~749798 and the ERC grant no.~757791, the National Natural Science Foundation of China (NSFC) under grant no.~61701155, the Fundamental Research Funds for the Central Universities under grant no.~JZ2019HGBZ0130, the Netherlands Organisation for Scientific Research (NWO) via the VIDI project ICONIC under grant no.~15685, and the Swedish Research Council under grant no.~2017-03702.

}

\newif\iffullbib

\fullbibfalse

\iffullbib

\bibliographystyle{IEEEtran}
\bibliography{references} 
\label{references}

\else

\end{document}